# Disorder-to-order transition of long fibers contained in evaporating sessile drops

S. Sannyamath[1,2], R. Vetter[3], H. Bonart[1], M. Hartmann[1], R. Ganguly[4], S. Hardt[1]*

[1]Institute for Nano- and Microfluidics, Department of Mechanical Engineering, TU Darmstadt, Germany
[2]Department of Civil Engineering, Jadavpur University, India
[3]Computational Physics for Engineering Materials, ETH Zürich, Switzerland
[4]Department of Power Engineering, Jadavpur University, India
*Corresponding author: S. Hardt, Institute for Nano- and Microfluidics, Department of Mechanical Engineering, TU Darmstadt, Peter-Grünberg-Str. 10 64287 Darmstadt, Germany. Email: hardt@nmf.tu-darmstadt.de

**Abstract**

A liquid drop containing a long fiber is a complex system whose configuration is determined by an interplay of elastic stresses in the fiber and capillary forces due to the liquid. We study the morphological evolution of fibers that are much longer than the drop diameter in evaporating sessile drops. After insertion, the fibers are either found in an ordered or disordered state, with increasing disorder for increasing fiber length. Upon evaporation, the order increases, in such a way that the final configuration deposited on the solid surface is either a circle, an ellipse, or 8-shaped. The morphology of the deposit depends on the fiber length and the elastocapillary length, both non-dimensionalized with the characteristic drop size, which we classify in a morphology regime map. The disorder-to-order transition allows depositing ordered fiber structures on solid surfaces even in cases of a strongly disordered state after fiber insertion. Combined with technologies such as inkjet printing, this process could open new avenues to decorate surfaces with filamental structures whose morphology can be controlled by varying the fiber length.

> **Significance**
>
> We demonstrate a self-organization process that allows depositing ordered filamental structures on solid surfaces. The process transforms disordered initial configurations of fibers contained in drops into ordered structures of three different types. Combined with technologies such as inkjet printing, this process could open new avenues to decorate solid surfaces with filamental structures.

# Introduction

Aggregation and deposition patterns of particles and fibers suspended in evaporating drops may either represent new perspectives or challenges for engineering applications, ranging from inkjet printing to bottom-up self-assembly of three-dimensional, mesoscopic and microscopic structures for micro-scale devices. Liquid droplet-mediated bottom-up self-organization of fibers on solid substrates has shown promise in a few applications such as passive alignment in micro-electronic and optical devices (1), fiber-scaffolds for tissue engineering (2) and textile engineering (3), and self-organizing micro- and mesoscale architectures (4). Especially in context of recent advances in additive manufacturing technology (AMT), the interaction of drops of complex liquids with surfaces has moved into the focus. Inkjet printing technologies (5), including recent versions such as electrohydrodynamic inkjet printing (6), form an important cornerstone of AMT, especially for complex liquids. Particularly, bioprinting applications highlight how complex the liquids to be deposited on a surface may become (7, 8). Once a drop has been deposited on a surface, it becomes important how the droplet-borne suspended or dissolved components distribute over the surface. An important but often undesired phenomenon in this context is the coffee-ring effect (9, 10) that leads to the deposition of dissolved or suspended components close to the three-phase contact line of an evaporating drop.

In that context, an especially complex scenario is a drop containing a single fiber much longer than the drop diameter. In nature, a similar phenomenon is found in glue droplets on spider silk (11). Silk threads can coil inside these droplets, and the result is a solid-liquid hybrid material with remarkable elastic properties. Another analogy from nature is the packaging of DNA in viral capsids (12), which is only possible in a multiply coiled configuration. There are proof-of-principle demonstrations of how drops interacting with deformable threads may enable novel technological applications, for example for stretchable electronics (13), or when soft objects need to be picked up from and released on a surface (14).

At least two different scenarios are possible when a multiply coiled fiber interacts with a drop. First, it may happen that the fiber encapsulates the drop, without penetrating into its interior. This will be referred to as the *fiber-on-drop* configuration. Second, a sufficiently long fiber contained in a drop will also coil, which will be referred to as the *fiber-in-drop* configuration. For both scenarios, the elastocapillary length

$$L_{ec} = \sqrt{\frac{Er^3}{\gamma}} \qquad (1)$$

is an important governing parameter (15), where $E$ is the elastic modulus of the fiber material, $r$ the radius of the fiber, and $\gamma$ the surface tension of the liquid. On a scale comparable to $L_{ec}$, the elastic bending forces in a fiber and the surface-tension forces of the liquid are of comparable magnitude. Therefore, a drop of radius $R$ will not experience any significant deformation by a fiber in the limit $R \gg L_{ec}$. In experiments with *fiber-on-drop* configurations (16), two different fiber conformations were observed, one where a fiber is only bent by a drop and a second where a fiber winds around a drop, even multiple times. Order-of-magnitude wise, the transition between the conformations occurs when the drop radius equals the elastocapillary length pertaining to the fiber and the droplet liquid.

For the case that a fiber has a strong affinity to a liquid, it can easily intrude into a drop, i.e., it is found in a *fiber-in-drop* configuration. The morphogenesis of semi-flexible polymers constrained on a non-compliant spherical surface was studied theoretically in the limit $R \gg L_{ec}$, which showed the existence of different forms of stable fiber-loops (16). The same limiting case was studied experimentally by introducing different fibers/wires in a rigid spherical cavity, and the influences of the friction between different fiber segments, the elasticity and plasticity of the fibers, and the torsion of the fibers were analyzed (17, 18). Similar studies, both experimental and simulations, were conducted for a case where the shell that encloses the fiber is deformable (19, 20). These studies revealed a complex interplay between the deformation of the fiber and that of the elastic shell. Depending upon different combinations of friction (between the contacting surfaces of filaments and the confinement) and the confinement rigidity, distinct morphological phases of long filaments (*spiral*, *classical*, *folded* and *warped*) were reported. The effect of friction in particular is complex, however; while pivotal in compliant shells (19, 20), it plays only a minor role in the packing of long fibers inserted into rigid spheres (17, 18, 30). A recent study has also focused on self-propelled fibers in spherical confinement, where the propulsion force acts tangentially on the fiber (21). It has been found that the propulsion promotes the transition from disordered to ordered fiber conformations.

In spite of these earlier works, one question appears to be largely unresolved: How can a drop containing a fiber or a filament induce a self-organization process that produces ordered structures of such one-dimensional materials? For example, it is not fully understood under which conditions and how the time-dependent confinement of a drop can affect the degree of order in the state of an immersed fiber. Related to that, a few studies for drops that are not in contact with a solid surface were conducted. In the theoretical/numerical work by Liu and Xu (22), one-dimensional nanomaterials (for example nanofibers) in a free evaporating droplet were considered, and it was demonstrated which morphological changes the nanofibers experience under increasing confinement. In the experiments conducted by Elettro et al. (23), a freely suspended *droplet-on-fiber* under tension was considered. Under moderate tension, the fiber coils inside the droplet, where ordered and disordered morphologies were observed. After the liquid has evaporated, either ring-shaped or 8-shaped morphologies were obtained. While these works have considered free-evaporating, fiber-laden droplets, to the best of our knowledge, fiber morphogenesis inside an evaporating sessile droplet has not been investigated. Such a study merits interest because the presence of a solid surface, as often encountered in practical applications, can have significant influence on the fiber behavior.

In our work, we study the evolution of fiber-drop systems on solid surfaces upon evaporation of the liquid. The experiments are performed with coiled fibers, i.e., fibers much longer than the initial drop diameter. The focus lies on the morphological changes of the fibers and their deposition on a surface. We show that the fiber deposits are well-ordered, where the morphology is determined by the non-dimensional fiber length and elastocapillary length. This opens up an avenue to create microstructured surfaces in a self-organization process.

## Results and Discussion

**Fiber morphogenesis in the droplet**

In a typical experiment, a droplet of purified and de-ionized water of specified volume (2.5, 5.0, 7.5 and 10 μL) is first dispensed on a superhydrophobic aluminum surface (*SI*, Fig S5A). The test section is adequately shielded to minimize external aerodynamic disturbances that would otherwise induce shear-induced advection inside the droplet. A single strand of viscose fiber of specified diameter (8 or 16 μm) and length (10, 25 or 50 mm) is then introduced in the sessile droplet by touching the liquid surface with the fiber. The strongly hydrophilic nature of the viscose fiber (*SI*, Fig. S5B) generates enough surface tension force to draw the fiber quickly into the droplet and also wrap the overhanging part of the fiber around the droplet (*SI*, Fig. S5C). In the confinement of the droplet, the fiber bends in conformity with the curvature of the droplet surface. Under the competing influence of capillary force, fiber elasticity, and interaction of the fiber with the underlying substrate, the fiber strand assumes a specific coil-morphology that varies with time as the droplet evaporates and shrinks. The fiber morphogenesis is recorded by imaging the droplet simultaneously from two orthogonal directions (top and side views, see Fig. 1A, and *SI*, Fig. S2).

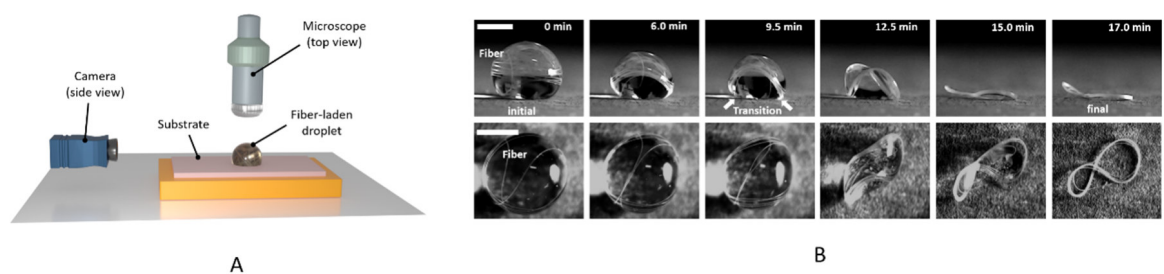

**Figure 1:** (A) Experimental setup for droplet imaging. (B) Image sequence in side view and top view of a 2.5 μl water droplet containing a 50 mm long and 16 μm diameter viscose fiber, showing the morphogenesis from an initial disordered configuration to a final 8-shaped depositional pattern via a saddle shaped anticlastic transition phase. Scale bars for both front and top view images denote 1.0 mm. The "transition" morphology at $t = 9.5$ min corresponds to the event of the fiber ends touching the substrate (see the arrows). The droplet evaporation time is 15 min while the image pair at 17 min corresponds to a completely dry state, displaying an 8-shaped fiber morphology.

Figure 1B shows an image sequence over a duration of 17 minutes while a 2.5 μL water droplet containing a 50 mm long and 16 μm diameter viscose fiber evaporates, leading to a progressively shrinking enclosure for the fiber (see the time-accelerated Movie 1 of the *SI*). The fiber exhibits a transition from an initially disordered (0 min) to a final 8-shaped, ordered deposition pattern (17.0 min) via an intermediate saddle-shaped anticlastic phase (12.5 min). Thus, from the top and side views in Fig. 1B, one may get the qualitative idea that the degree of order of the fiber increases progressively as the encapsulating droplet shrinks. A more quantitative description of the degree of order is provided in the following section. During the entire phase of evaporation of the droplet, particularly at the later stages (e.g., beyond $t = 9.5$ min for Fig. 1B), the coiled fiber morphology is found to have a significant influence on the droplet shape, i.e., the latter deviates noticeably from the shape of a droplet without fiber. This warping of the droplet surface is a result of the competition between elastic stresses in the fiber due to its coiling, and capillary stresses due to the curved liquid surface.

**Initial fiber morphology: Simulation and experiment**

To get a deeper insight into the elastocapillary buckling of the fibers, particularly with respect to how the initial fiber morphology emerges upon introduction of the fiber inside the liquid confinement, numerical simulations, based on the classical beam theory for uniform, isotropic, linear elastic media are performed (24). The fiber is injected into the droplets at constant speed at a specified angle of insertion α, measured between the fiber and a tangent plane to the droplet surface at the point of insertion. The time evolution of the fiber inside the droplet is computed by solving Newton's equations of motion with a predictor-corrector integration method with adaptive time-step control, using nodal forces and moments derived from the internal and external energy functionals, steric repulsion, and friction. Notably, in the simulations the presence of the solid surface is not accounted for, which means that the simulation model is only able to make predictions for scenarios in which the fiber does not get in touch with the surface. Specific details of the simulation are elaborated in Appendix F.

To quantify the different observed fiber configurations upon complete fiber insertion into the droplet, we use a morphological order parameter $f$ that measures the degree of flattening of the coiled fiber strand as,

$$f = \frac{a-b}{a} \tag{2}$$

with $a = \sqrt{\lambda_1}$ and $b = \sqrt{\lambda_3}$ where $\lambda_1 \geq \lambda_2 \geq \lambda_3$ are the eigenvalues of the 3×3 covariance matrix $C = \int_0^l \vec{x}(s)\, \vec{x}(s)^T\, ds$ of the fiber's centerline $\vec{x}(s)$, $s \in [0, l]$. The flattening can be equivalently expressed as

$$f = 1 - \frac{1}{\sqrt{\kappa(C)}} \tag{3}$$

where $\kappa(C)$ is the condition number of $C$ in the 2-norm. The order parameter of the fiber is maximal ($f = 1$) if the fiber is flat, such as a planar ring or ellipse, and nearly maximal also for an 8-shape (Fig. 2A). For a fiber with unit aspect ratio $a/b$, such as a disordered spherical arrangement, the order parameter is minimal ($f = 0$).

Unlike in the simulation, it was not possible to have a precise control over the insertion angle of the fiber strand in the experiments because the fibers were either too flexible (to have a straight overhung portion beyond the point they were held by the tweezers) or they had pre-existing bends and curls due to residual stress in the dry fibers. To account for this variation of the insertion parameters in a simplified manner, we have simulated the fiber insertion into the spherical droplet for different angles of insertion α (see Fig. 2B and C). Interestingly, the simulations reveal that the insertion angle strongly affects the degree of order in the morphology of the fully immersed fiber. For radially directed insertion ($\alpha = 90°$), the fiber primarily coils into an ordered torus perpendicular to the insertion axis, which minimizes its curvature and elastic energy, largely independent of its length and elastocapillary length (Fig. 1B, see Movie 2 of the *SI*). Only rarely, some disorder is observed. This large degree of order results from the fact that the parts of the fiber that have already merged into the coil come to a rest and do not move further as more fiber is being inserted and coiling up. Only the recently immersed parts of the fiber move and merge into the circular fiber bundle. Fiber-fiber friction and fluid drag, therefore, do not contribute significantly to the loading process when the fiber is injected radially.

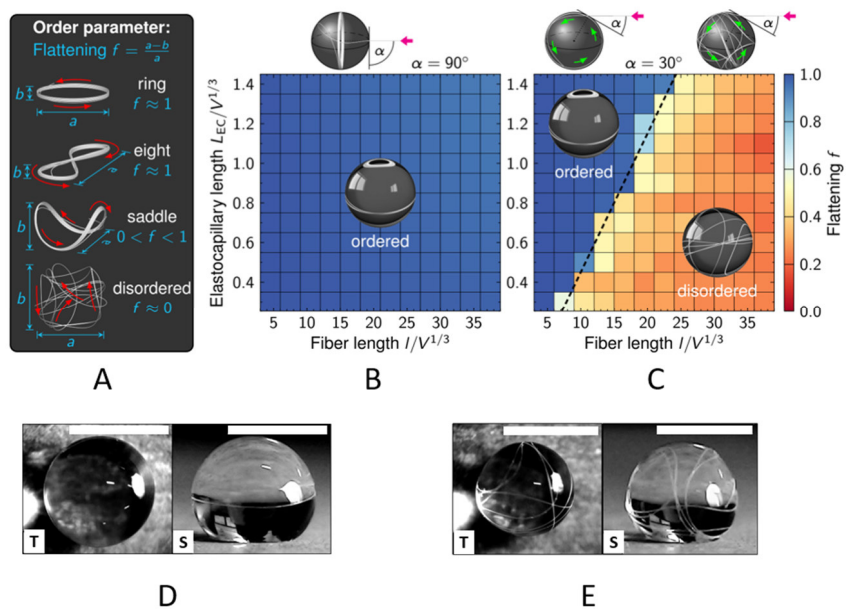

**Figure 2:** Initial morphology of a fiber in a droplet. (A) Sample fiber configurations and their corresponding order parameter $f$. Red arrows indicate the fiber orientation, cyan arrows show the maximal and minimal square-root eigenvalues of the covariance matrix, $a$ and $b$. Morphological phase diagram, as obtained from the simulations and plotted on a normalized elastocapillary length - fiber length plane for (B) centered injection ($\alpha = 90°$), and (C) nearly tangential injection ($\alpha = 30°$). Both the lengthscales are normalized by the characteristic droplet size, $V^{1/3}$. The phase boundary is indicated with a black dashed line. Magenta arrows point along the injection axis at angle $\alpha$ to the water-air interface, green arrows indicate bulk rotation of the immersed part of the fiber. Each colored square represents the median order parameter over 10 independent finite element simulations. Insets show raytraced images of exemplary fiber-laden droplets from the respective morphological phases, with a refractive index of 4/3 at the water-air interface. Images from experiments of ordered (D) and disordered (E) fiber morphologies (T: top view, S: side view) after fiber insertions in a 7.5 μL droplet. The scale bar denotes 2.0 mm. The fiber length is 100 mm for (D) and 50 mm for (E). The fiber diameter is 16 μm in both the cases.

With more tangential insertion (i.e., smaller $\alpha$), this changes dramatically: The fiber initially coils up in the great circle that lies in the plane spanned by the insertion axis and the droplet center, and the entire bulk coil rotates, pushed along by the continued fiber injection. For short fibers, this ordered morphology remains present until the entire fiber is drawn inside the droplet (Movie 3 of the *SI*). However, as the fiber-insertion continues, the drag force between the liquid and the fiber (proportional to the immersed fiber's surface area) and inertial force (proportional to the immersed fiber's volume) build up with the immersed length of the fiber. Eventually, the force required to feed in more fiber exceeds the force required to buckle the fiber out of the coiling plane. Once the coiling symmetry is broken in this way, disordered loading ensues, with frequent changes in the coiling direction, which reduce the flattening $f$. The fiber undergoes rigid-body rotation inside the droplet, without much relative rearrangement (Movie 4 of the *SI*). Thus, when we plot the value of $f$ on a normalized elastocapillary length-fiber length plane, a sharp order-disorder transition is observed (Fig. 1C). At $\alpha = 30°$, the morphological phase boundary is well approximated by a straight line $L_{EC}/l = 0.07$ (the black dashed line in Fig. 1C). Thicker fibers (greater $L_{EC}$) and shorter fibers (smaller $l$) coil up into an ordered configuration even for nearly tangential insertion, because the axial compression from drag, inertia and friction is not large enough to buckle them. Thinner fibers (smaller $L_{EC}$) and longer ones (greater $l$) buckle out of the coiling plane more easily, which gives rise to the disordered phase. For more centered fiber insertion (e.g., above $\alpha \approx 60°$), the ordered phase begins to grow until it occupies nearly the entire phase space at $\alpha = 90°$. Similar order-disorder transitions have previously been found for elastic fibers confined in spheres and flexible shells (18, 19, 20), but with intrinsic fiber curvature, friction and the flexibility of the shell identified as free parameters along which the transition occurs, rather than the fiber injection angle and fiber length as reported here.

These simulations were carried out taking into consideration the fiber-fiber friction (see Appendix F). A very similar morphological diagram, exhibiting an order-to-disorder transition at the same parameter location, but with somewhat more flattening in the disordered phase, is found with frictionless fibers (*SI*, Fig. S1). This is in agreement with previous work that reports only small effects of friction when the spatial confinement of the fiber does not deform substantially (17, 18, 30). Moreover, we quantified the degree of (dis)order in the fiber arrangements also with two other order parameters, the average coiling direction $|P|$ (*SI*, Fig. S2) and a local definition of the classical nematic order parameter $S$ (*SI*, Fig. S3). While both are not equivalent to the flattening $f$ (*SI*, Fig. S4), the order-to-disorder transition is found at the same location in the examined parameter range for all three definitions of (dis)order, corroborating the robustness of the transition with respect to different interpretations of fiber order.

The experimental runs with fibers of different lengths and diameters and droplets of different volumes show a similar morphology upon fiber insertion, i.e., ordered for the smaller fiber lengths, and disordered for the longer ones. This can be seen in Fig. 4A, which shows a morphology map of the initial fiber configuration in a space spanned by the dimensionless fiber length and dimensionless elastocapillary length. In the experiments, the initial fiber morphology was determined through visual inspection. A configuration in which a fiber is largely arranged in one plane was considered ordered, whereas for the case that no single plane could be identified with which the strands of a fiber are aligned, a disordered morphology was assigned. In the ordered morphology, the fibers remain coiled with its strands bundled together, predominantly on its equatorial plane. This allows the fiber to remain coiled within the droplet with the largest possible radius of bend, thus minimizing the elastic energy. On the other hand, in a disordered morphology, the fiber is found to be loosely coiled in different planes, crisscrossing itself. Nonetheless, even in this configuration, the fiber resides near the droplet surface to entail minimum elastocapillary bending (25). Unlike in the simulations, the dependence of the phase boundary between order and disordered domains on the insertion angle could not be assessed from the experimental runs. This may be attributed to the limited control over the fiber insertion process in the experiments. There is a fundamental difference in the fiber insertion mechanism in the simulation vis-à-vis during the experiment. In the simulation, we assign a constant insertion velocity of the fiber, penetrating the liquid surface. On the contrary, the insertion in the experiment predominantly occurs tangentially (see Movie 5 of the *SI*), with surface tension spooling the fiber into the droplet at an unspecified speed. Apart from the capillary-driven spooling, other forms of fiber insertion are also observed in the experiment. For example, after one end of the fiber is drawn in the droplet, the overhanging fiber simply warps on the droplet surface in some occasions. In some cases, the overhanging portion of the fiber touches the substrate first, then drags/pulls along the surface and finally coils into the droplet (see Movie 6 of the *SI*).

While the experiments and the simulations are in approximate agreement, it is worth noting that an increase of the fiber length has the opposite effect with respect to the fiber morphology compared to what was reported in (23). In (23), it was found that the coiling of fibers under tensile stress contained in droplets is characterized by an increasing order when the fiber length relative to the droplet diameter increases, and when the droplet is significantly deformed due to the presence of the fiber. Neither is the latter criterion fulfilled in our experiments, nor is a tensile stress applied to the fibers, which prevents a direct transfer between the different sets of experiments. Still, these results show that the physics of long fibers inside drops is complex and that more work is required to get an overview of the complete configuration space.

**Fiber morphogenesis as the droplet evaporates**

Having identified the initial fiber morphologies, we now characterize how the fiber morphology evolves as the droplet evaporates. A specific case of morphological transition of the fiber inside the gradually shrinking confinement of the droplet is presented in Fig. 1B, where the fiber is found to evolve from an initially disordered configuration to a final ordered one. As can be seen in Fig. 1B, a transition occurs where the fiber gets in touch with the substrate. Since the simulations described in the previous section did not take into account the presence of a solid surface, we are able to study the fiber morphogenesis only experimentally. By contrast, usually the fiber insertion in the droplet takes place without contact between the fiber and the solid surface, and hence we were able to compare the fiber behavior upon insertion both from the simulations and the experiments. Overall, the temporal evolution of a fiber during droplet evaporation appears smooth, albeit certain discontinuities are observed in specific cases. For example, Fig. 3 shows the fiber morphogenesis in a 5 µL droplet in which a 50 mm long, 16 µm diameter fiber initially exhibits a disordered morphology (Fig. 3A). The time evolution remails quasi-static (Fig. 3A through D) as the droplet evaporates, i.e., it occurs on a timescale large compared to the intrinsic timescales of the system, until a sudden "flipping" is observed (Fig. 3E, Movie 7 of the *SI*) with a rapid change in fiber orientation. After this sudden event, the fiber morphology again evolves in a quasi-static manner. While the exact reason of the "flipping" remains unknown, we suspect that this event arises as the coiled fiber attempts to reduce its compressive stress arising due to droplet shrinkage, and plays a key role in transforming a disordered configuration to an ordered one. In all the experiments, an increasing order during the time evolution of droplets was observed. Even initial configurations with pronounced disorder transformed into ordered fiber deposits on the surface. The final stages of the time evolution of a fiber-drop system, including the fiber deposits on the surface, are documented in Figs. S7, S8 and S9 of the *SI*. Figs. S7 and S9 show two scenarios leading to 8-shaped fiber deposit. Once the fiber is in the saddle configuration and the droplet evaporates further, it becomes energetically favorable to twist the fiber bundle rather than bend it further, so the saddle collapses into an 8.

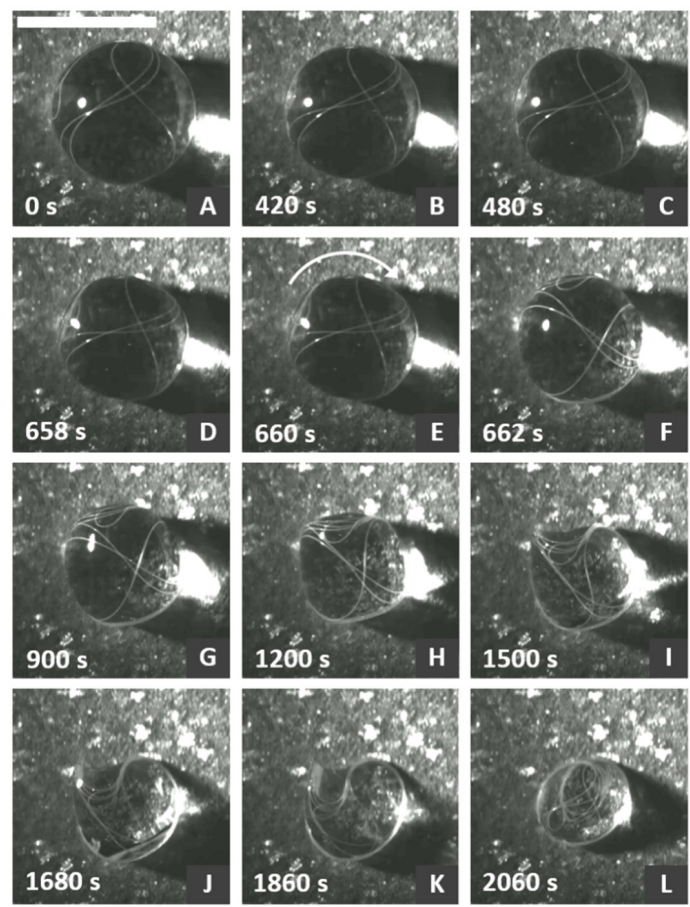

**Figure 3.** Morphogenesis of a 50 mm long, 16 μm diameter fiber inside a 5 μL water droplet from an initial disordered to a final 8-like structure. The morphological transformation is accompanied by a sudden flipping of a section of the fiber, visible in frame E. The arrow indicates the direction of fiber movement during flipping. The scale bar denotes 2.0 mm.

**Morphogenic regime map**

To develop a morphogenic regime map that represents the initial, intermediate and final fiber configurations, we have performed experiments with different droplet volumes (2.5, 5.0, 7.5 and 10 μL), fiber diameters (8 and 16 μm) and fiber lengths (10, 25 and 50 mm). The parametric regimes are summarized in Table S1 of the *SI*. Fiber morphologies are identified by analyzing the top and side views captured throughout the period of droplet evaporation. These morphologies for the initial (right after the fiber insertion in the droplet is complete), intermediate (when the fiber takes an anticlastic saddle morphology, with its two legs touching the base of the droplet; albeit, for the circular intermediate morphology no noticeable difference from the initial ordered structure is observed), and final (when the droplet liquid has completely evaporated) conditions are plotted in Fig. 4 on a plane spanned by the elastocapillary length ($L_{EC}$) and the fiber length ($l$), both normalized by the droplet lengthscale ($V^{\frac{1}{3}}$, where $V$ is the initial droplet volume). Each of the 24 data points in the plots represents a summary of 5 repetitions, where the frequency of a specific morphology is described through a pie chart on each symbol. The background color represents the probabilities of the morphologies classified via Gaussian processes conditioned on the experimental data points, see Appendix E.

The initial fiber morphology in Fig. 4A tends to be ordered for short fibers, i.e., for $lV^{-\frac{1}{3}} \lesssim 20$, while longer fibers exhibit a disordered morphology. As already mentioned, this is in qualitative agreement with the numerical simulations reported in Fig. 2. Deviations between the experiments and the simulations are attributed to the limited control of the insertion process in the experiments, as discussed above.

A comparison of Figs. 4A and B indicates that for $lV^{-\frac{1}{3}} \lesssim 15$, the initial fiber morphology of a circular ring residing at the equatorial plane of the droplet is highly probable. Beyond $lV^{-\frac{1}{3}} \approx 15$, the fibers tend to buckle under the compressive stress of the shrinking confinement. Despite buckling, the fibers conform with the surface of the shrinking and deforming droplets, engendering a saddle-shaped morphology. In most cases, the lower ends of the saddle are found to touch the substrate at the droplet base, where they remain pinned during the remaining period of droplet evaporation (Fig. 1B (iii, iv)). In the saddle morphology, a significant departure of the droplet shape from the shape without fiber (well-approximated by a spherical cap) is seen (insets iii and iv of Fig 4B).

The final fiber morphology can be classified via three distinct shapes, viz., circular (see insets i and ii of Fig. 4C, where the difference between the diameters along the principal axes is less than 10%), elliptic (insets iii and iv of Fig. 4C, where this difference is more than 10%), and 8-shape (insets v and vi of Fig. 4C). A comparison of Figs. 4B and C shows that for fibers with $V^{-\frac{1}{3}} \lesssim 10$, i.e., fibers that have an ordered initial and a circular intermediate morphology, it is extremely likely that eventually a circular fiber deposit is found on the substrate. Longer fibers that start with an ordered morphology tend to end up in an elliptic final morphology via a saddle-shaped intermediate regime. For fibers that start with a disordered morphology, the probability is very high to yield an 8-shaped deposit on the substrate via the saddle-shaped intermediate morphology. A careful observation of the circular, elliptic and 8-shaped fibers reveal that the circular and elliptic deposits are essentially planar, with their number of turns increasing with the fiber length. In contrast, 8-shaped fibers are sometimes not in an entirely two-dimensional configuration, as segments of the fibers occasionally warp upward (e.g., in Fig. 1Bvi).

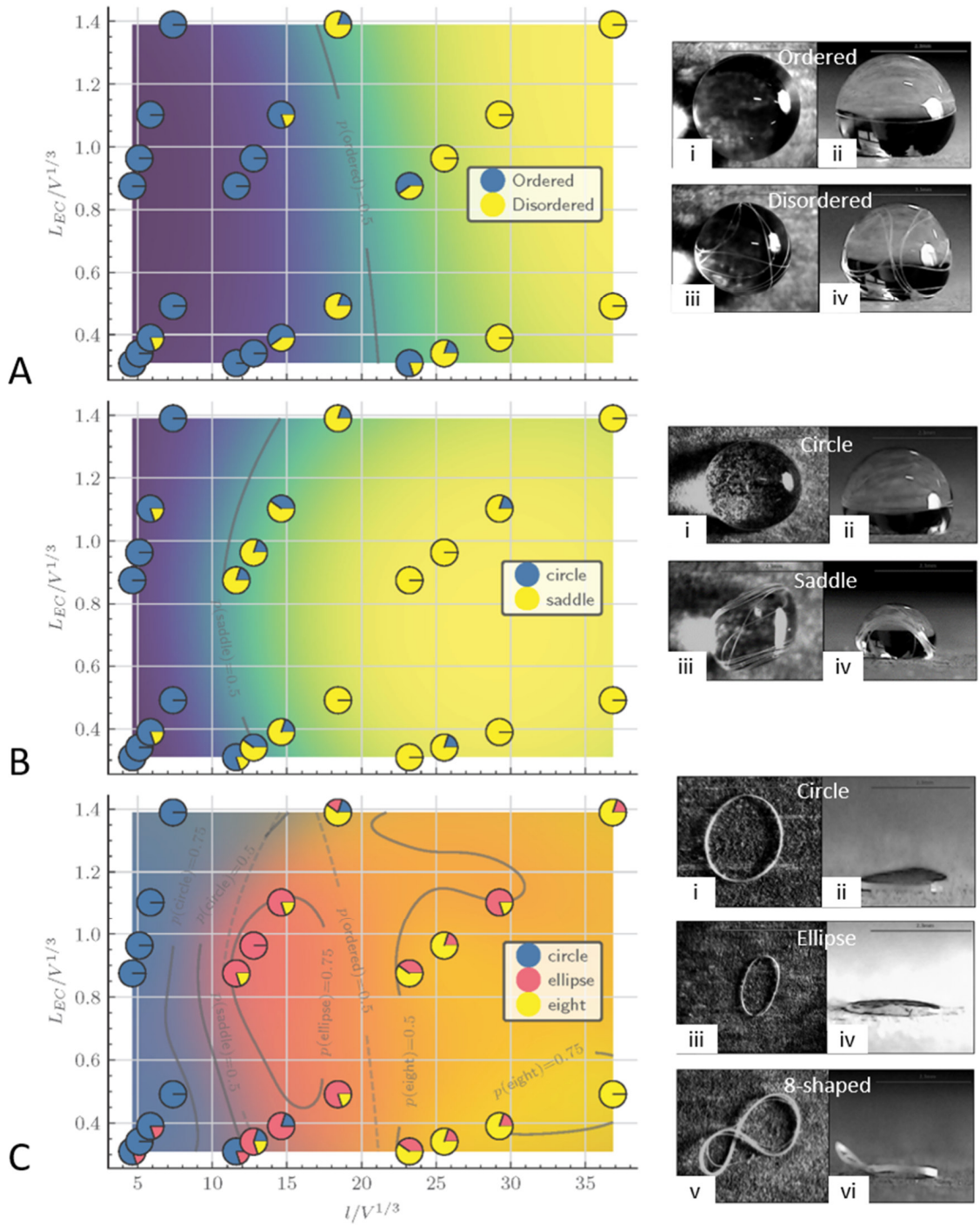

**Figure 4.** Fiber morphology regime map on a plane defined by the elastocapillary length ($L_{EC}$) and the fiber length ($l$), both normalized by the droplet lengthscale ($V^{1/3}$): (A) initial, (B) intermediate and (C) final morphology. Each of the 24 cases indicated in the regime maps corresponds to five independent experimental runs, and the results are shown as pie-charts. The background colors are the morphology probabilities from Gaussian process classification. As a droplet evaporates, the fiber contained in a droplet gets reconfigured. It passes through a circular or saddle-shaped morphology at the intermediate stage (B), eventually settling on the substrate in a circular, elliptical or 8-shaped morphology (C).

Just as the droplet evaporation dynamics influences the fiber morphogenesis, the presence of fiber also influences the droplet evaporation rate. As the droplet surface gets warped in response to the elastic stress in the fiber, particularly at the later phase of dry-out, the droplet shape significantly deviates from a spherical cap (*SI,* Fig. S10A). Warping of the liquid surface increases the surface-to-volume ratio, among others, leading to an increased evaporation rate. A parameter study shows that the droplet dry-out time reduces with increasing fiber volume fraction (*SI,* Fig. S10B). An increased fiber volume fraction distorts the droplet shape from the spherical cap shape, thus increasing its effective surface-to-volume ratio, which in turn enhances the evaporation rate. For the parametric range reported in Table S1 of the *SI*, the droplet lifetime $\tau$ (expressed in min) is found to scale with the fiber volume fraction $\varphi$ (ratio of the fiber volume to the droplet volume at the time of fiber insertion) as $\tau \approx 42.4 e^{-0.80\varphi}$.

## Conclusions

To conclude, we have studied the morphogenesis of hydrophilic elastic fibers contained in an evaporating sessile water droplet on a superhydrophobic surface. The most remarkable aspect of the presented results is the fact that an evaporating droplet serves as a progressively shrinking enclosure of the fiber, creating order from a disordered state. A fiber of a disordered, three-dimensional initial configuration ends up either as an ordered, two-dimensional circular, elliptical or 8-shaped deposit on the surface. Additional research is needed to fully understand the emergence of the different morphologies during droplet evaporation.

The process we have studied allows creating ordered structures of quasi-one-dimensional materials on solid surfaces in a self-organization process. Combined with technologies such as inkjet printing, this could open new avenues to decorate surfaces with filamental structures whose morphology can be controlled by varying the fiber length. Our research can be extended in different directions, for example by considering variations of the surface wettability or the evaporation of fiber-laden drops on surfaces with a topography.

# Appendix

## A: Surface fabrication and functionalization

The superhydrophobic aluminum surfaces were fabricated following a protocol described in (26), with appropriate modifications for the desired surface wettability. First, aluminum plates (procured from Ernst Wenzelmann Schilderfabrik GmbH, grade Al-6061) of dimensions 40mm × 40mm × 2mm were cleansed by detergent to remove any oil or dirt. Then the plates were mechanically grinded and roughened with sandpaper (granulations 220, 400, 800, 1000, and 2000, respectively) and polished in a polishing unit (Chennai Metco, PMV009) to remove the native oxide film on the surface. Subsequently, the plates were thoroughly cleaned by sequential rinsing with acetone, ethanol, isopropyl alcohol (IPA, Sigma Aldrich) and finally, deionised (DI) water. The cleansed samples were immersed in 2.0 M aqueous hydrochloric acid (HCl, Merck) solution for 15-20 minutes. Subsequently, they were removed from the HCl solution and immersed in ultrasonic baths (USB 3.5L H DTC, PCI Analytics) of acetone and distilled water for 30 minutes each to remove residual imperfections, if any. Finally, the samples were air-dried and preserved at room temperature for 30 min. The chemically etched surfaces exhibit a strong super-hydrophilic (SHPL) nature and hemiwicking (27). The reactions involved in the wet chemical etching are:

$$2\,Al\,(s) + 6\,HCl\,(aq) \rightarrow 2\,AlCl_3\,(aq) + 3\,H_2\,(g) \tag{4a}$$

$$Al - 3\,e^- \rightarrow Al^{3+} \tag{4b}$$

$$2\,H^+ + 2\,e^- \rightarrow H_2 \tag{4c}$$

$$2\,Al + 6\,H^+ + 6\,Cl^- \rightarrow 2\,Al^{3+} + 6\,Cl^- + 3\,H_2 \tag{4d}$$

The abbreviated form of final redox reaction may be summarized as:

$$2\,Al + 6\,H^+ \rightarrow 2\,Al^{3+} + 3\,H_2 \tag{4e}$$

The etched samples were further boiled in deionised water to passivate the surface, facilitating the growth of hierarchical micro-nano structures of a thin layer of aluminum oxide hydroxide (AlO(OH)), more commonly known as böhmite. To render the surfaces superhydrophobic (SHPB), the etched plates were immersed in 1.0% (wt./vol.) pure ethanolic (99.8%) solution of 1H,1H,2H,2H-Perfluorooctyltriethoxysilane (also termed fluoroalkylsilane, or FAS, procured from Alfa Aesar) for $1^1/_2 - 2$ h. Finally, the plates were cured in an air-oven at 100 ºC for 2 h, and preserved carefully for future use. This procedure allows grafting of a self-assembled monolayer of FAS onto the surfaces. The low surface energy of FAS, coupled with the hierarchical roughness of the surface leads to the desired level of superhydrophobicity.

## B: Surface characterization

In all experiments, deionized water (Milli-Q) water was used. In order to characterize the wettability of the SHPB samples, systematic measurements of dynamic contact angles were made using a contact angle measurement apparatus (Holmarc Opto-Mechatronics Pvt. Ltd, HTCAM58). The fabricated superhydrophobic surfaces exhibited an advancing contact angle of $CA_{adv}$ =167.5 ± 0.7º and a receding contact angle of $CA_{rec}$=164.1 ± 1.1º, implying a very low contact angle hysteresis ($CAH = CA_{adv} - CA_{rec}$) of ~4º (see *SI*, Fig. S5A). Topographic features of the surface were examined with an optical microscope (Magnus Zoom MSZTR with MagCam DC-5 camera). The superhydrophobic aluminum surfaces have an RMS roughness value of $Ra$ = 2.65 µm, obtained using an Atomic Force Microscope (Asylum Research GmbH).

**C: Properties of viscose fibers**

Elastic viscose fibers (procured from Kelheim Fibres GmbH), having a density of $\rho$ = 0.83 g/cm³ (28) and moisture content of ~ 11-14% at 65% relative humidity, were chosen for insertion into the water droplets. The cross-section of the fibers is not exactly circular but has an undulated boundary. Therefore, the reported fiber diameters are approximate values that are obtained by replacing the cross-sectional shape by a circle of the same area. The Young's modulus, measured using an Atomic Force Microscope (Asylum Research GmbH), is $E_{fiber}$ = 450 ± 50 MPa. These measurements were carried out with fibers immersed in water. The fibers show negligible shrinkage in water (28). Fig. S5B of the *SI* shows the water affinity of the fibers: the contact angle on the fiber surface almost vanishes, and fibers are rapidly sucked into the liquid, indicating their superhydrophilic nature.

**D: Experimental setup**

The experimental setup is depicted in Fig. 1A (see also *SI*, Fig. S6). The superhydrophobic substrates were placed on a vibration-free horizontal platform. Water droplets were carefully dispensed onto the surfaces using a mechanical pipette (Eppendorf Research Plus, 2-12 µL), followed by the manual introduction of viscose fibers with the help of tweezers. Contacting the droplet with a fiber triggered its rapid suction into the droplet owing to the high wettability of the fiber. A stereo zoom optical microscope (Magnus Zoom MSZTR with MagCam DC-5 camera, connected to a desktop computer), was employed for imaging the fiber-laden droplets from the top. Depending on the droplet size, images were captured with 1.0x, 1.2x, 1.5x, or 2x objective magnifications, while maintaining a pixel aspect ratio of 1:1 for the field of view. A Micro-ware 1000x USB digital microscope (Mmpl-1000x microscope, Microware Multimedia Pvt. Ltd.), connected to a laptop was used simultaneously for the side view image acquisition. Cross-illumination, perpendicular to the optical axes of both top and side view imaging, was provided using a cold light source (Western Surgical Alpha Dual, 250W), with its intensity set to ~30%. Experiments were performed with different combinations of droplet volumes (viz., 2.5, 5.0, 7.5, and 10.0 µL), fiber diameters (8 and 16 µm), and fiber lengths (10, 25, and 50 mm). All experiments were conducted under controlled ambient conditions of $T_{air}$ = 25 ± 0.5 ºC, relative humidity $\phi$ = 50 ± 5%, and for each case a series of five experiments was performed.

**E: Data evaluation**

The description of ordered vis-à-vis disordered fiber morphology from the experimental images is based on its visual attribute. If the majority of the fiber strand is oriented along a single plane inside the droplet, we term it ordered. On the contrary, a disordered fiber strand would have a major part of it intertwined within the entire droplet, crisscrossing itself several times in different planes.

To determine the probabilistic regime maps and extract general trends, we classify the experimental points using Gaussian processes as available from scikit-learn 1.2.2 (29) with an isotropic radial basic function (RBF or squared exponential) kernel and default parameters. This nonparametric supervised learning method is widely used in probabilistic classification problems. Given the experimental data points, GPC provides a predictive distribution over possible functions that could explain the data. Kernels, or covariance functions, compute how similar two points are. The RBF kernel is easily applied due to its similarity to the well-known Gaussian distribution and does not make any strong assumptions about the data. As input features we choose the elastocapillary length and the fiber length, while the output is the state of the fiber as determined from the images. Each feature is scaled individually to range between 0 and 1. The hyperparameters of the kernel are then estimated from the data. We train three different GPCs; one for every state of the experiment (initial, intermediate, final). Afterwards we can use the trained GPCs to calculate the probabilities of each category (ordered/disordered, circle/saddle, circle/ellipse/eight) for every given elastocapillary length and fiber length. The resulting classification probabilities are visualized using Matplotlib 3.7.1 (30). The accuracy score calculated with scikit-learn is above 80% for all panels in Fig. 4.

**F: Finite element simulations**

We model the fiber with classical beam theory based on uniform, isotropic, linear elasticity, as described in detail in (24). In brief, the fiber's internal state of stress is expressed by the energy functional

$$U_{\text{internal}}[\varepsilon, \kappa, \varphi] = \frac{1}{2}\int_0^l ds \left( EA\varepsilon(s)^2 + EI\kappa(s)^2 + \frac{EI}{1+\nu}\varphi'(s)^2 \right) \quad (5)$$

where $l$ is the unstretched fiber length, $A = \pi r^2$ the cross-sectional area with radius $r$, $I = \pi r^4/4$ the second moment of area, $E$ Young's modulus, and $\nu$ the Poisson ratio. The first term in the integrand models elastic tension or compression, using the axial Cauchy strain $\varepsilon(s) = \|\vec{x}'(s)\| - 1$, where $\vec{x}(s)$ traces the fiber's centerline, parameterized by the undeformed arclength $s$. The second term accounts for bending, with centerline curvature $\kappa(s) = \|\vec{x}''(s)\|$. Finally, the orientation of the fiber's cross section about its centerline is encoded in the twist angle $\varphi(s)$, whose rate of change along the arc-length is penalized by the third energy term, the twisting energy. The fiber is discretized into finite elements, and the displacement field is approximated with cubic Hermite splines on each element, as is done routinely in finite element analysis of beams. We set the elastic moduli to $E = 500$ MPa and $\nu = 1/3$.

We simulate the fiber insertion into spherical water droplets of radius $R$ with fibers discretized into elements of length of $0.08R$. The fiber is constrained to a straight feeding axis as long as it is still outside the droplet, and this constraint is lifted for fiber segments that have already been immersed. The water-air interface is represented by a deformable triangulated mesh governed by a surface tension of $\gamma = 72$ mN/m; at the location of fiber insertion the bending moment (approximately $EI/R$) of the fibers is found to deform the water droplets only by a small fraction of the fiber diameter. We therefore simplify the liquid-air interface to a rigid sphere for the parameter screens, and use its surface tension value only to compute the elastocapillary length. Confinement into the droplet is modeled with a frictionless steric potential that penalizes penetrations of the fiber into the liquid-air interface of the droplet. Formally, these two constraints can be written as an external energy functional

$$U_{\text{external}}[\vec{x}] = \frac{k}{2} \int_0^l c(\vec{x}(s)) ds, \tag{6}$$

where $k$ is a Lagrange multiplier chosen in the order of the fiber's Young's modulus. The integrand is set to

$$c(\vec{x}(s)) = y(s)^2 + z(s)^2 \tag{7}$$

to keep the fiber on the feeding axis as long as it has not entered the droplet yet ($\|\vec{x}(s)\| > R - r$), and

$$c(\vec{x}(s)) = (\|\vec{x}(s)\| - R + r)^2 H(\|\vec{x}(s)\| - R + r) \tag{8}$$

to keep it inside the droplet once it has entered it. $H$ denotes the Heaviside step function, and $\vec{x}(s) = [x(s), y(s), z(s)]^T$ is the fiber's parameterized centerline position.

To account for contact of the fiber with itself, repelling Hertzian normal forces are made to act when two fiber segments touch, and a slip-stick Coulomb friction model is implemented for relative tangential movements of touching fiber segments, as described earlier (30). The friction model is parameterized by two coefficients, the static and dynamic Coulomb coefficients $\mu_s = 1$ and $\mu_d = 0.9$.

The fiber is injected into the droplets at constant speed. We solve Newton's equations of motion with a predictor-corrector integration method with adaptive timestep control, using nodal forces and moments derived from the internal and external energy $U = U_{\text{internal}} + U_{\text{external}}$, steric repulsion, and friction. As we do not resolve the fluid motion in the droplet explicitly, we additionally apply subcritical viscous damping to the translational and rotational movement of the immersed fiber, mimicking the drag exerted by the fluid on the fiber. Once the entire fiber is immersed, the system is evolved further for a short time period until static mechanical equilibrium is reached.

### List of Movies

1. Movie 1: experiment, time-evolution of the fiber-in-drop system corresponding to Fig. 1B, accelerated 30 times;
2. Movie 2: simulation, fiber insertion at an angle of 90º, ordered morphology;

3. Movie 3: simulation, fiber insertion at an angle of 30º, disordered morphology;
4. Movie 4: simulation, fiber insertion at an angle of 30º, ordered morphology;
5. Movie 5: experiment, fiber insertion in predominantly tangential mode;
6. Movie 6: experiment, fiber insertion while the overhanging fiber is dragged over the substrate;
7. Movie 7: experiment, sudden "flipping" event.

## Author Contributions

S.S. performed research and analyzed data, R.V. performed research, contributed new analytic tools, analyzed data and wrote the paper, H.B. contributed new analytic tools and analyzed data, M.H. performed research, R.G. designed research, analyzed data and wrote the paper, S.H. designed research, analyzed data and wrote the paper.

## Author's competing interest and contact information

The authors declare no competing interests.

**Email addresses**:

S.S.: shatadru2014@gmail.com
R.V.: vetterro@ethz.ch
H.B.: bonart@nmf.tu-darmstadt.de
M.H.: maximilian.hartmann@focused-energy.world
R.G.: ranjan.ganguly@jadavpuruniversity.in
S.H.: hardt@nmf.tu-darmstadt.de

## Data sharing plan

The research data corresponding to this work will be made available via TU Darmstadt's TUdatalib repository (https://tudatalib.ulb.tu-darmstadt.de/)

## Acknowledgements

The authors acknowledge the support by Matthias Kühnhammer, Soft Matter at Interfaces, TU Darmstadt, who performed the AFM measurements. S.S. acknowledges the support by the German Academic Exchange Service (DAAD), grant number 57460839/ 91691578. R.V. acknowledges the support by ETH Zürich, grant number ETH-03 10-3. H.B. acknowledges the support by the Deutsche Forschungsgemeinschaft (DFG, German Research Foundation) – project IDs 459970814; 459970841. M.H. and S.H. acknowledge the support by DFG within the Collaborative Research Centre 1194 ''Interaction of Transport and Wetting Processes'' Project-ID 265191195, subproject A02b.

Supplementary Information to

# Disorder-to-order transition of long fibers contained in evaporating sessile drops

S. Sannyamath[1,2], R. Vetter[3], H. Bonart[1], M. Hartmann[1], R. Ganguly[4], S. Hardt[1]*

[1]Institute for Nano- and Microfluidics, Department of Mechanical Engineering, TU Darmstadt, Germany
[2]Department of Civil Engineering, Jadavpur University, India
[3]Computational Physics for Engineering Materials, ETH Zürich, Switzerland
[4]Department of Power Engineering, Jadavpur University, India
*Corresponding author: S. Hardt, Institute for Nano- and Microfluidics, Department of Mechanical Engineering, TU Darmstadt, Peter-Grünberg-Str. 10 64287 Darmstadt, Germany. Email: hardt@nmf.tu-darmstadt.de

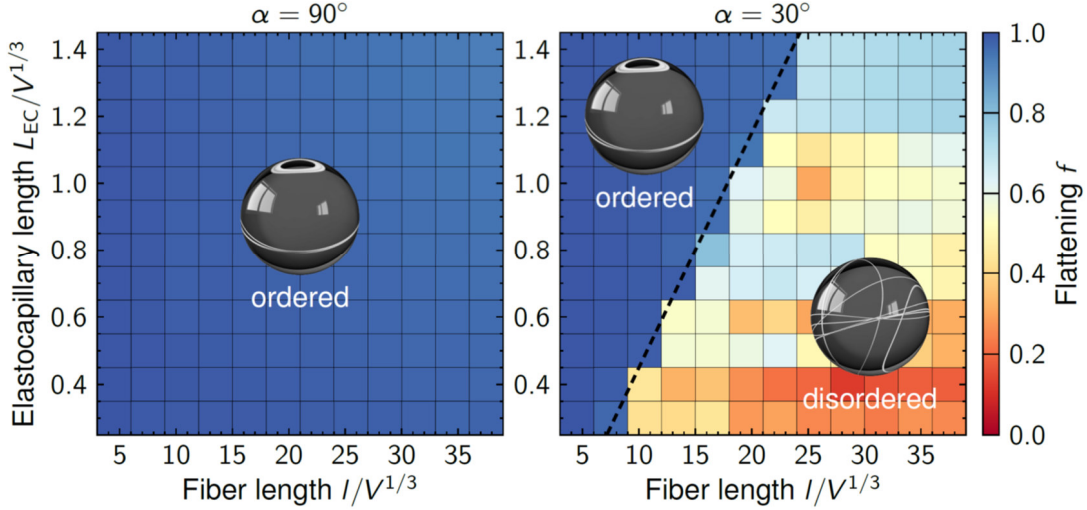

**Figure S1.** Morphological phase diagram analogous to Fig. 2B,C, but without considering fiber-fiber friction. See the caption of Fig. 2 for details.

As alternative measures of fiber order, we introduce two more morphological order parameters. The first one is closely related to the one originally proposed to describe order-disorder transitions of fiber packings in flexible shell confinements (Ref. 20 of the main text), but more robust to perturbations away from perfect fiber alignment. It measures the unsigned mean coiling direction of the fiber about its own principal axis of maximal moment of inertia, and is defined as follows. Denoting the fiber's centerline centroid by

$$\vec{c} = \langle \vec{x} \rangle = \frac{1}{l}\int_0^l \vec{x}(s)\, ds \qquad (S1)$$

and by $\vec{v}_3$ the eigenvector with the largest eigenvalue of its 3×3 inertia tensor $I$ about $\vec{c}$, whose components are given by

$$I_{ij} = \int_0^l \left( \|\vec{x}(s) - \vec{c}\|^2 \delta_{ij} - (\vec{x}(s) - \vec{c})_i (\vec{x}(s) - \vec{c})_j \right) ds, \quad i,j = 1,2,3, \qquad (S2)$$

$\vec{v}_3$ points along the fiber's principal axis of maximal moment of inertia. For a torus, this is always the axis of its rotational symmetry. The triple product

$$\varpi(s) = (\vec{v}_3 \times \vec{x}'(s)) \cdot (\vec{x}(s) - \vec{c}) \qquad (S3)$$

is negative if the fiber turns right around the line running along $\vec{v}_3$ and passing through $\vec{c}$, and positive if it turns left around it. To measure the degree of order in the coil, the sign of this quantity is averaged over the fiber length

$$P = \langle \mathrm{sgn}(\varpi) \rangle = \frac{1}{l} \int_0^l \mathrm{sgn}(\varpi(s))\, ds. \qquad (S4)$$

$P$ vanishes if half of the fiber turns left and the other half right. For a circular coil, $P = \pm 1$, whereas for an isotropic, disordered fiber arrangement, $\vec{v}_3$ points in a random direction and $P = 0$. Since mirror-symmetric fiber configurations are considered equivalent here, we use as order parameter the coiling order magnitude $|P| \in [0,1]$. 8-shaped fibers have half of their segments turning one way and half the other, implying $|P| = 0$, whereas unidirectional ring-like coils (perfectly circular, oval, or mild perturbations of such shapes such as a saddle shape) have $|P| = 1$ (Fig. S2, left).

The phase diagram that uses $|P|$ to distinguish between ordered and disordered fiber configurations (Fig. S2) is qualitatively identical to the one using the flattening $f$ (Fig. 2), but the disordered phase is somewhat more heterogeneous in terms of $|P|$.

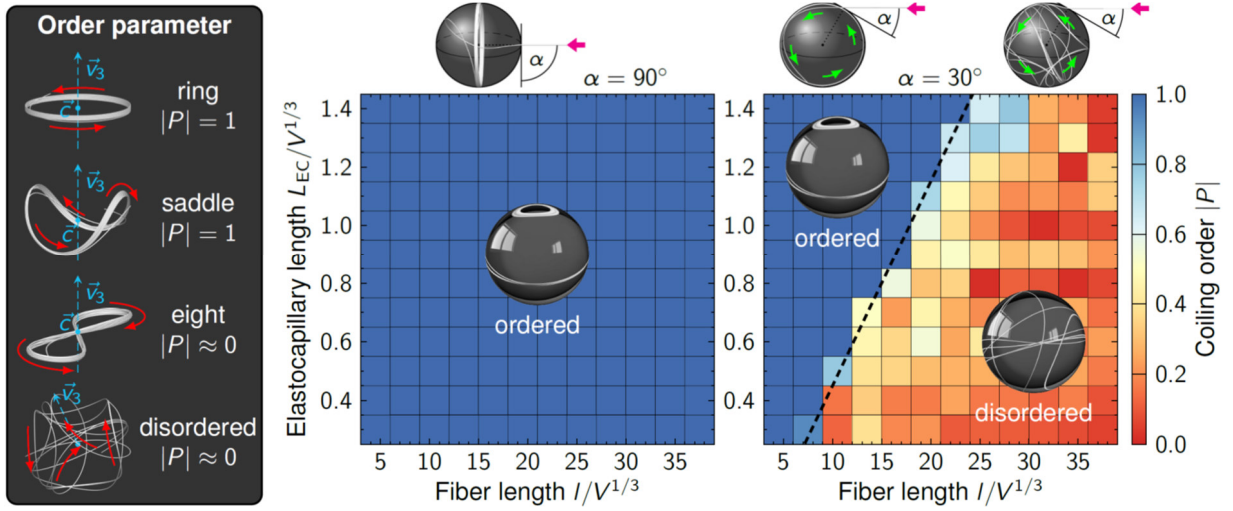

**Figure S2.** Order-to-disorder transition in numerical simulations analogous to Fig. 2, using the coiling order $|P|$ as the order parameter. Each colored square represents the median order parameter over 10 independent finite-element simulations.

As a third way to quantify fiber order, we borrow the notion of nematic order from liquid crystal theory and evaluate it in a local sense. Intuitively, the order in the fiber arrangement is high if adjacent fiber segments tend to be parallel, and low if they are randomly oriented or preferentially orthogonal. This is captured by the following local definition of the classical nematic order parameter $S$:

$$S = \langle \frac{3\cos^2\theta - 1}{2} \rangle = \frac{3}{2l} \int_0^l \cos^2\theta(s, a)\, ds - \frac{1}{2}, \tag{S5}$$

where $\theta(s, a)$ is the mean angle between the fiber tangent at the arclength position $s$ and other tangents within a local neighborhood of radius $a$. We use $a = 5r$ as the neighborhood size here, where $r$ is the fiber radius, as this includes up to next-nearest neighboring fiber segments, but excludes all other segments that are further away. Distance is measured from the fiber centerline.

We observe the local nematic order $S$ to produce a qualitatively identical phase diagram for the initial fiber configurations (Fig. S3) as $f$ (Fig. 2) and $|P|$ (Fig. S2), notably with the same phase boundary.

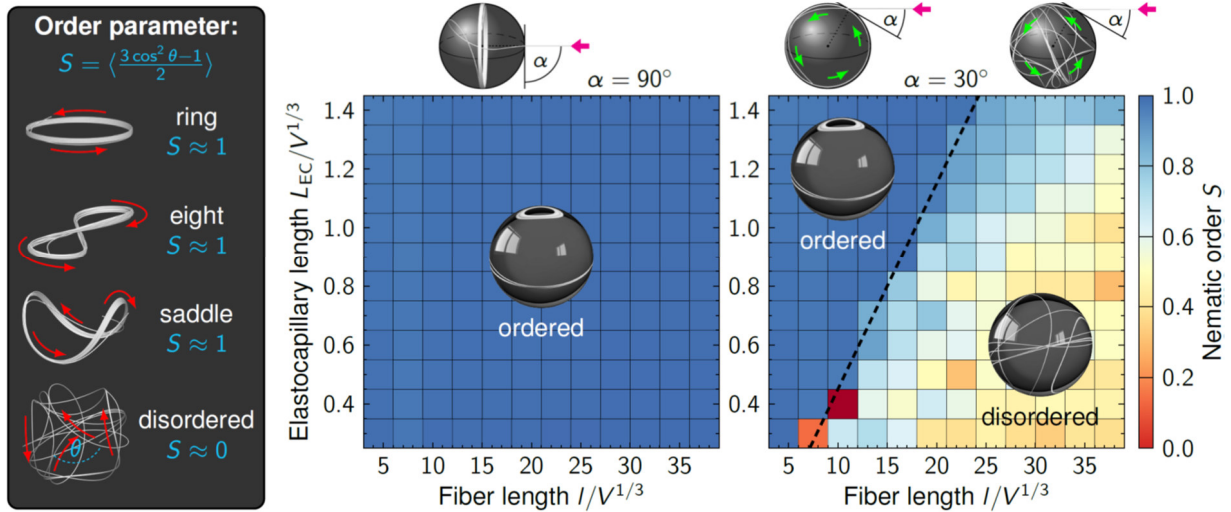

**Figure S3.** Order-to-disorder transition in numerical simulations analogous to Fig. 2, using the nematic order $S$ as order parameter. The order is evaluated locally, within a neighborhood (center-to-center distance) of $5r$ within each fiber segment, where $r$ is the fiber radius. Each colored square represents the median order parameter over 10 independent finite element simulations.

Although the order-to-disorder transition occurs for all three order parameters at the same independent system parameters, their mutual relationships are nonlinear. Figure S4 shows scatterplots for each pair of order parameters. While the flattening is never near zero for a fiber longer than $2\pi R$, the coiling order can take values arbitrarily close to zero. A coiling order of $|P| = 1$ tends to be found in 2D-like flat configurations (large $f$), but in particular for long fibers, can also be found in more three-dimensional fiber arrangements. On the other hand, a high degree of nematic order also tends to occur in flat fiber configurations, but is also found when the coiling order is relatively low. Finally, in rare cases with relatively short fibers, we observe negative nematic order ($S < 0$) due to nearly orthogonal fiber contacts.

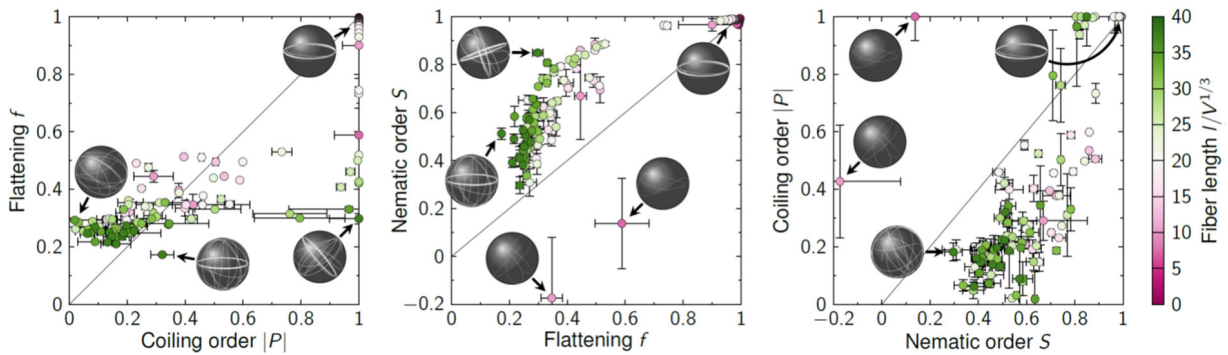

**Figure S4.** Pairwise comparison of different order parameters. Data points and error bars represent medians and their standard errors from 10 independent finite element simulations. Inset images show exemplary configurations to illustrate particular cases.

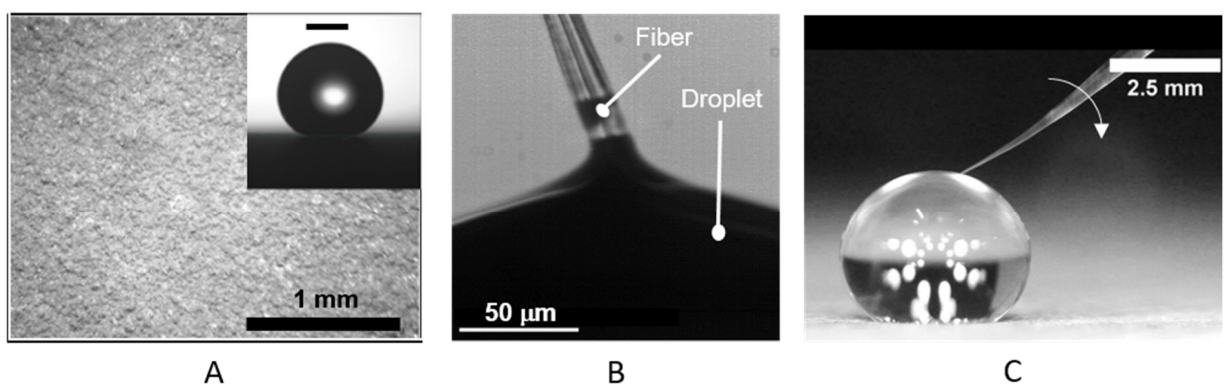

**Figure S5.** (A) Optical microscopy image of the superhydrophobic aluminum surface. The inset shows a water droplet on the surface. The scale bar in the inset represents 1 mm. (B) Micrograph of fiber insertion into a droplet. The liquid meniscus at the fiber indicates a contact angle ~ 0°. (C) Snapshot of a representative fiber insertion event. The overhanging part of the fiber is observed to warp around (indicated by the white arrow) the droplet immediately after contacting the liquid surface.

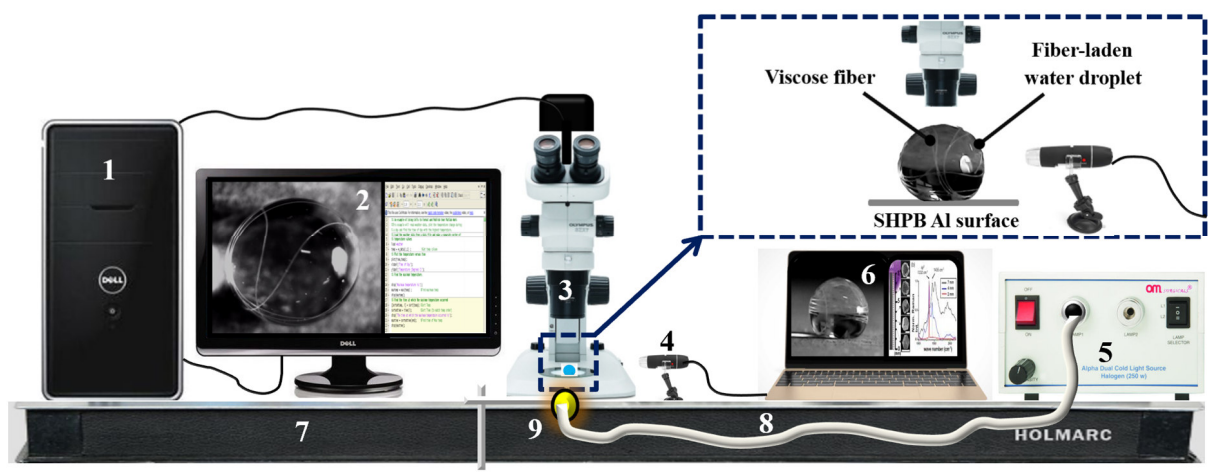

**Figure S6.** Schematic diagram of the experimental setup for studying the evaporation of fiber-laden droplets on superhydrophobic aluminum surfaces. Legend: 1: Desktop computer; 2: Screen-feed of the top view of a fiber-laden droplet; 3: Stereo zoom digital microscope for top view image acquisition; 4: Micro-ware 1000x USB digital microscope for side view image acquisition; 5: Cold light source for cross illumination; 6: Screen-feed of the side view of a fiber-laden droplet; 7: Optical breadboard; 8: Optical fiber for cold light; 9: Front end goose neck of the cold light source.

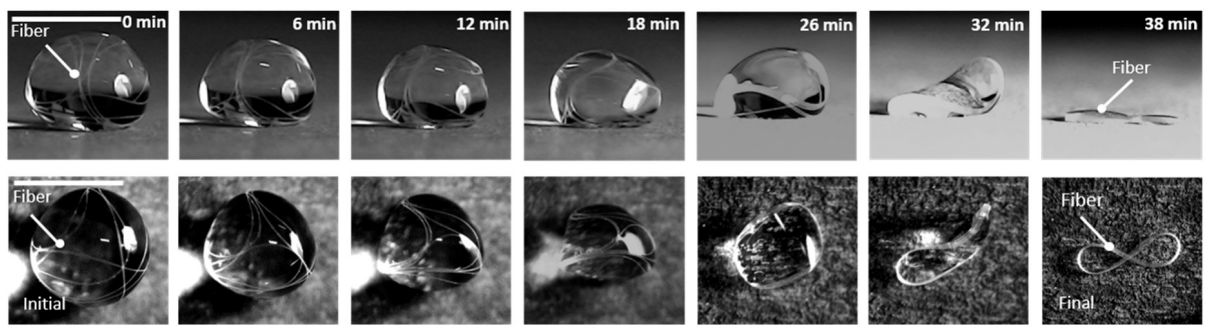

**Figure S7.** Time-lapse images of a 7.5 μL water droplet laden with a 50 mm long and 8 μm diameter viscose fiber, showing the morphogenesis from an initial disordered configuration to a final 8-shaped depositional pattern via a saddle-shaped transition phase. Scale bars for both side and top view images indicate 2.0 mm. The droplet evaporation time is 38 min. Sudden flipping of the coiled fiber is observed between 12 and 18 minutes.

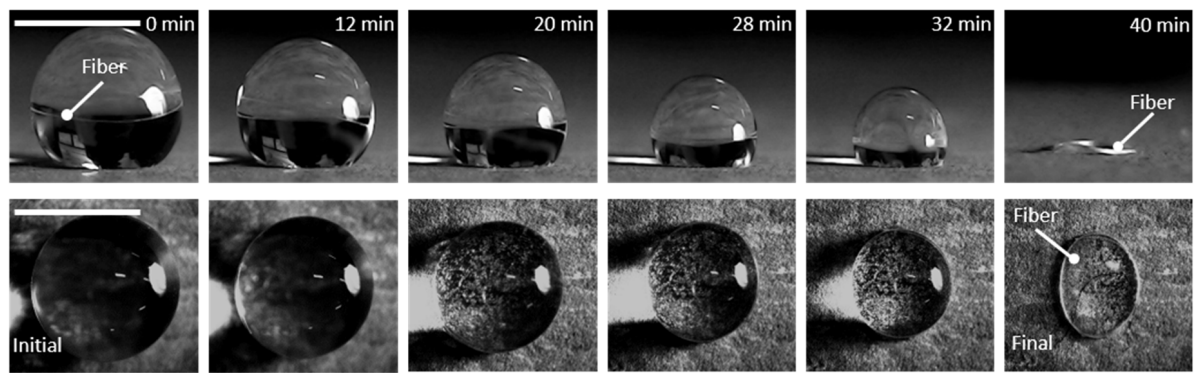

**Figure S8.** Time-lapse images of a 7.5 μL water droplet laden with a 10 mm long and 16 μm diameter viscose fiber, showing the morphogenesis from an initial ordered configuration to a final circle-shaped depositional pattern. The scale bars for both side and top view images indicate 2.0 mm. The droplet evaporation time is 40 min.

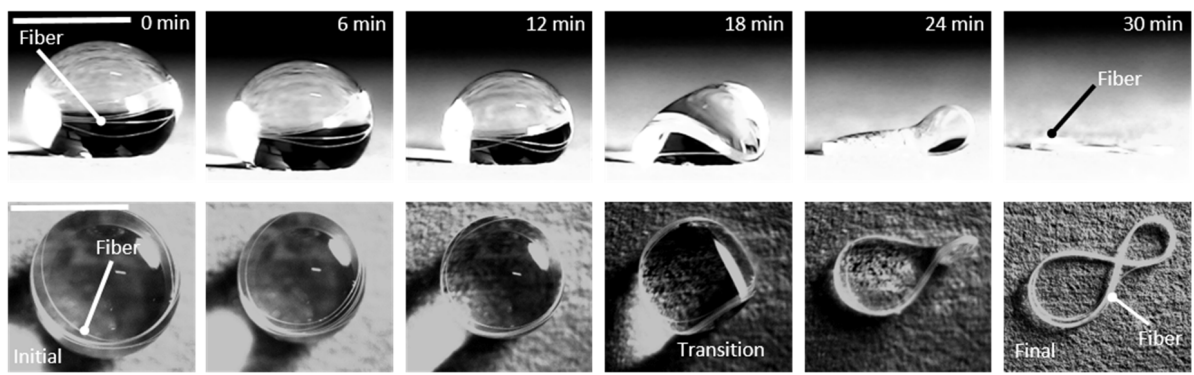

**Figure S9.** Time-lapse images of a 10 μL water droplet laden with a 50 mm long and 16 μm diameter viscose fiber, showing the morphogenesis from an initial ordered configuration to a final 8-shaped depositional pattern via a saddle-shaped intermediate state (occurring at ~18 minutes, marked as "transition", when the coiled fiber touches the substrate at two positions). The scale bars for both side and top view images denote 2.0 mm. The droplet evaporation time is 30 min, i.e., shorter than the evaporation time of the smaller droplets shown in Figs. S3 and S4.

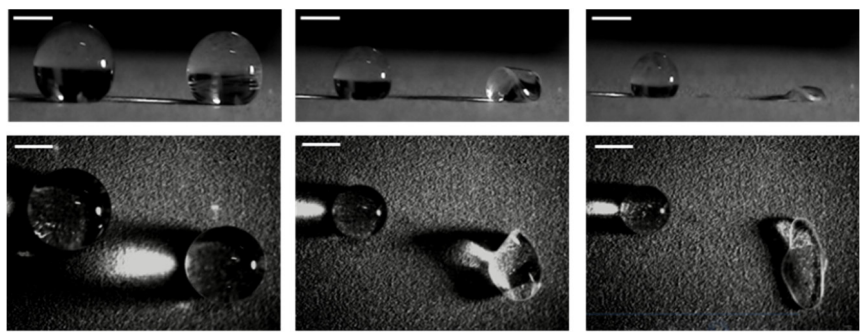

A

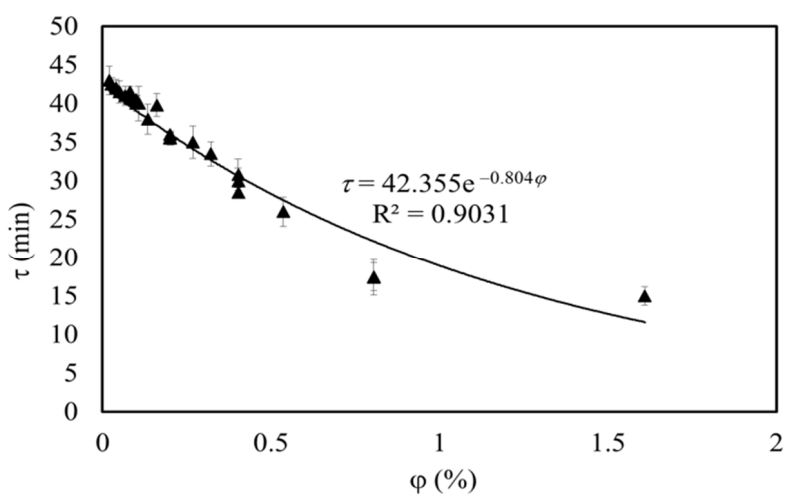

B

**Figure S10.** (A) Comparison of the evaporation dynamics of DI water droplets, without (left) and with a fiber (right), on a superhydrophobic aluminum surface. A representative initial droplet volume of 5.0 μL is chosen. The droplet on the right contains a 16 μm diameter and 25 mm long viscose fiber. The morphogenesis of the fiber shows a transition from an initial ordered to a final elliptical depositional pattern via an intermediate saddle morphology. The scale bars in the top and side views indicate 1 mm. (B) Plots describing the droplet lifetime $\tau$ as function of the initial fiber volume fraction $\varphi$ (= fiber volume / initial droplet volume) for different combinations of fiber diameter (8 and 16 μm), fiber length (10, 25 and 50 mm) and initial droplet volumes (2.5, 5, 7.5 and 10 μL). The symbols represent experimental data, with error bars indicating standard deviations of droplet lifetimes for each set of five experiments. The line represents an exponential fit to the data. The results indicate that even a small fiber volume fraction of the order of 1 % or less has a substantial effect on the droplet lifetime.

**Table S1: Parameters of the experiments**

| Experiment No. | Droplet volume $V$ (µL) | Fiber diameter $d$ (µm) | Fiber length $l$ (mm) | Experiment No. | Droplet volume $V$ (µL) | Fiber diameter $d$ (µm) | Fiber length $l$ (mm) |
|---|---|---|---|---|---|---|---|
| 1 | 2.5 | 8 | 50 | 13 | 7.5 | 8 | 50 |
| 2 | | | 25 | 14 | | | 25 |
| 3 | | | 10 | 15 | | | 10 |
| 4 | 2.5 | 16 | 50 | 16 | 7.5 | 16 | 50 |
| 5 | | | 25 | 17 | | | 25 |
| 6 | | | 10 | 18 | | | 10 |
| 7 | 5 | 8 | 50 | 19 | 10 | 8 | 50 |
| 8 | | | 25 | 20 | | | 25 |
| 9 | | | 10 | 21 | | | 10 |
| 10 | 5 | 16 | 50 | 22 | 10 | 16 | 50 |
| 11 | | | 25 | 23 | | | 25 |
| 12 | | | 10 | 24 | | | 10 |